\documentclass[12pt]{article}
\usepackage{amsmath}
\usepackage{epsfig}
\usepackage{amsfonts}

 \title{Depurification by Lorentz boosts}
 \author{ Piotr Kosi\'nski\thanks{supported by the {\L}\'od\'z University grant N$^0$ 795.},
 Pawe{\l}  Ma\'slanka$^*$ \\
Department of Theoretical Physics II \\
University of {\L}\'od\'z \\
Pomorska 149/153, 90 - 236 {\L}\'od\'z/Poland.}
 \date{}
 \begin{document}
 \maketitle
\begin{abstract}
We consider a particle of half-integer spin which is
nonrelativistic in the rest frame. Assuming the particle is
completely polarized along third axis we calculate the Bloch
vector as seen by a moving observer. The result for its length is
expressed in terms of dispersion of some vector operator linear in
momentum. The relation with the localization properties is
discussed.
\end{abstract} \newpage
As it is well known the quantum information processing gets
modified if special relativity is taken into account \cite{b1}
$\div$\ \cite{b8}. This can be demonstrated even in the simplest
case of free relativistic particle with spin \cite{b2}. Contrary
to the Galilean case where boosts commute, the commutator of
Lorentz boosts gives rotation; this property underlies, for
example, Thomas precession. As a result, spin variable transforms
in momentum - dependent way and this mixing between spin and
momentum degrees of freedom results in depurification of reduced
spin density matrix \cite{b2}, \cite{b4}. The amount of
depurification can be related to the localization properties of
the state \cite{b4}.

In the recent paper \cite{b9} we have derived the general formula
for spin density matrix for moving observer. We have shown that
under suitable conditions Lorentz boosts can transform the pure
spin state into totally unpolarized one.

In the present note we use the result of \cite{b9} to find the
Bloch vector in the moving reference frame for totally polarized
particle which is nonrelativistic in the rest frame. We find also
the formula for depurification in terms of dispersion of the
observable linear in momentum. This result confirms the general
relation between depurification and localization.

Assume we have a nonrelativistic spin $-\frac{1}{2}$\ particle in
such a pure state that the reduced spin density matrix describes
pure spin state. Then, by a suitable rotation, one can arrange
things such that $a(p,2)=0$. The momentum wave function $a(p,1)$\
is in turn supported in the region $\mid \vec{p}\mid \ll m$. The
spin density matrix $\rho$\ is completely determined by Bloch
vector
\begin{eqnarray}
\vec{\mu}=Tr(\rho \vec{\sigma}) \label{w1}
\end{eqnarray}
The assumption made above implies
\begin{eqnarray}
\vec{\mu}=\vec{e}_3\label{w2} \end{eqnarray} in the rest frame. We
want to calculate Bloch vector after applying the boost
characterized by the fourvelocity
\begin{eqnarray}
u^{\mu}=(ch\beta,\;sh\beta\vec{n}),\;\;\;\mid \vec{n}\mid
=1\label{w3}
\end{eqnarray}
It has been shown in Ref. \cite{b9} that the transformed Bloch
vector reads
\begin{eqnarray}
&&\vec{\mu}(\beta )=\int\frac{d^3\vec{p}}{2p_0}(cos\Omega \mid
a(p,1)\mid^2\vec{e}_3+sin\Omega (\vec{e}_3\times \vec{e})\mid
a(p,1)\mid^2 +\nonumber \\
&&+(1-cos\Omega )\vec{e}(\vec{e}\cdot \vec{e}_3)\mid
a(p,1)\mid^2)\label{w4}
\end{eqnarray}
where
\begin{eqnarray}
&&\vec{e}\equiv \frac{\vec{p}\times \vec{n}}{\mid \vec{p}\times
\vec{n}\mid} \nonumber \\
&&sin\Omega
=2\frac{\left((1+u^0)(p^0+m)-\vec{u}\cdot\vec{p}\right)\mid
\vec{p}\times \vec{u}\mid}{\left((1+u^0)(p^0+m)-\vec{u}\cdot
\vec{p}\right)^2+\mid \vec{p}\times \vec{u}\mid^2} \label{w5}\\
&&cos\Omega
=\frac{\left((1+u^0)(p^0+m)-\vec{u}\cdot\vec{p}\right)^2-\mid
\vec{p}\times \vec{u}\mid^2}{\left((1+u^0)(p^0+m)-\vec{u}\cdot
\vec{p}\right)^2+\mid \vec{p}\times \vec{u}\mid^2}\nonumber
\end{eqnarray}
Now, taking into account that $\mid a(p,1)\mid^2$\ is supported in
the region $\mid \vec{p}\mid \ll m$\ we can expand $sin\Omega$\
and $cos\Omega$\ in powers of $\frac{\mid\vec{p}\mid}{m}$. One
obtains
\begin{eqnarray}
&&sin\Omega \simeq \frac{\lambda}{m}\mid \vec{p}\times \vec{n}\mid
(1+\frac{\lambda}{2m}\vec{p}\cdot \vec{n}) \nonumber \\
&&cos\Omega \simeq 1-\frac{\lambda^2}{2m^2}(\vec{p}\times
\vec{n})^2\label{w6} \\
&&\lambda \equiv \sqrt{\frac{u^0-1}{u^0+1}}\nonumber
\end{eqnarray}

Inserting (\ref{w6}) into eq. (\ref{w4}) we arrive at the
following formula for transformed Bloch vector
\begin{eqnarray}
&&\vec{\mu}(\beta )=\left(1-\frac{\lambda^2}{2m^2}<(\vec{p}\times
\vec{n})^2>\right)\vec{e}_3+\frac{\lambda}{m}\vec{e}_3\times
(<\vec{p}>\times \vec{n})+\nonumber \\
&&+\frac{\lambda^2}{2m^2}<(\vec{p}\times \vec{n})((\vec{p}\times
\vec{n})\vec{e}_3)+(\vec{e}_3\times (\vec{p}\times
\vec{n}))(\vec{p}\cdot \vec{n})> \label{w7}
\end{eqnarray}
Eq. (\ref{w7}) provides the general expression for the Bloch
vector for the nonrelativistic (in the rest frame) particle as
seen by moving observer. In order to check whether
$\vec{\mu}(\beta )$\ describes pure or mixed state we take the
square of eq. (\ref{w7}) keeping terms up to the order
$\frac{\vec{p}^2}{m^2}$. The result can be described as follows.
For any vector operator $\vec{z}$\ define
\begin{eqnarray}
(\Delta \vec{z})^2\equiv \sum\limits_{i=1}^{3}(\Delta
z_i)^2=\sum\limits_{i=1}^{3}<(z_i^2-<z_i>^2)> \label{w8}
\end{eqnarray}
and let
\begin{eqnarray}\vec{\Pi}\equiv \vec{p}\times
\vec{n}-(\vec{e}_3(\vec{p}\times \vec{n}))\vec{e}_3\equiv
(\vec{p}\times \vec{n})_{\perp}\label{w9}
\end{eqnarray}
Then we get
\begin{eqnarray}
\mid \vec{\mu}(\beta )\mid^2=1-\frac{\lambda^2}{m^2}(\Delta
\vec{\Pi})^2\label{w10}
\end{eqnarray}

Note that the $\vec{\Pi}$\ is a linear combination of momentum
variables; therefore, there exists no normalizable state such that
$(\Delta \vec{\Pi})^2=0$. The transformed state is necessarily a
mixed one.

In the small velocity limit $\lambda^2\simeq \frac{\vec{v}^2}{4}$,
where $\vec{v}$\ is the observer velocity, and eq. (\ref{w10})
reads
\begin{eqnarray}
\mid \vec{\mu}(\beta )^2=1-\frac{\vec{v}^2}{4m^2}(\Delta
\vec{\Pi})^2\label{w11}
\end{eqnarray}
On the other hand, the extreme relativistic limit $\lambda =1$\
and
\begin{eqnarray}\mid \vec{\mu}(\beta )\mid^2=1-\frac{(\Delta
\vec{\Pi})^2}{m^2}\label{w12}
\end{eqnarray}
Eq. (\ref{w10}) allows to relate the amount of depurification to
the localization properties of the initial state. It is well known
\cite{b10} that there is no unique notion of position operator in
relativistic quantum theory. The reason is that any casual and
covariant interaction breaks necessarily the particle number
conservation law. However, we assumed that, in the rest frame, the
particle is nonrelativistic and the position operator is
well-defined. Moreover, $\vec{\Pi}$\ is linear in momentum which
suggest that the formula (\ref{w10}) can be expressed in terms of
localization of our particle. Following standard derivation of
uncertainty relations it is not difficult to show that
\begin{eqnarray}
(\Delta \vec{\Pi})^2(\Delta \vec{x})^2\geq \frac{1}{4}\label{w13}
\end{eqnarray}
and eq. (\ref{w10}) implies the following inequality
\begin{eqnarray}
\mid \vec{\mu}(\beta )\mid^2\leq 1-\frac{\lambda^2}{4m^2(\Delta
\vec{x})^2}\label{w14}
\end{eqnarray}
One should here keep in mind that we are in the nonrelativistic
regime and the particle is localized in region much larger then
Compton wavelength, $(\Delta \vec{x})^2\gg \frac{1}{m^2}$.

\end{document}